\documentstyle[12pt,aps,epsfig,array,preprint]{revtex}
\begin{document}

\title{Tuning the scattering length on  the ground 
triplet state of Cs$_2$.}
\author{V. Kokoouline, J. Vala, and R. Kosloff }
\address{Fritz Haber Research Center for Molecular Dynamics, 
Hebrew University of Jerusalem, Jerusalem, 91904, Israel}
\date{\today}
\maketitle
\begin{abstract}

We develop two schemes for tuning the scattering length on the ground 
triplet state of Cs$_2$. First, an absolute value of the  triplet 
scattering length of $^{133}$Cs$_2$ is determined using the experimental 
data (Fioretti {\it et al}, Eur.Phys.J. {\bf 5}, 389 (1999)).
We demonstrate that the large scattering length  can be made small and 
positive by coupling of the  $^3\Sigma_u^+ (6S + 6S)$ potential to
the $^3\Pi_g$ state by strong off-resonant radiation.
A weaker laser field coupling the $^3\Sigma_u^+ (6S + 6S)$  
continuum to the lowest bound level of the excited  
$^3\Sigma_g^+ (6S + 6P)$ also leads to a small positive scattering length.
In addition, the scattering length of the $^{135}$Cs isotope is found 
to be positive. The method used solves
the Schr\"odinger equation for two electronic states coupled by 
an electromagnetic field with approximations employed. 
The scattering length is determined from calculated continuum 
wavefunctions of low energy.

\end{abstract}
%\pacs{34.50.Rk ; 32.80.Qk ; 34.20.Cf} 

\newpage
A positive scattering length is a crucial ingredient in obtaining
Bose-Einstein condensation from a gas of ultra-cold atoms.
This parameter controls the long range repulsion between the atoms
and therefore the stability of the process.
Manipulating the scattering length will allow to add new candidates to the 
small list  of atomic systems which form a condensate.
Schemes to control the scattering length have been suggested based 
on manipulating the inter-atomic interactions by external fields. 
They include the use of a magnetic field to induce a Feshbach resonance 
\cite{tiesinga92,tiesinga93,vogels97,inouye98,courteille98,vuletic99a,vuletic99b}, 
a use of radio-frequency \cite{moerdijk96} and dc electric fields\cite{marinescu98} 
and off resonant strong electromagnetic field \cite{fedichev96,bohn97,kagan97}.
In the present study we propose two schemes for modifying the 
scattering length on the ground triplet state of Cs$_2$ using
electromagnetic field. First, 
we calculate the scattering length $a_{sc}$ of triplet $^{133}$Cs$_2$ and discuss the sensitivity of $a_{sc}$ to the $C_6$ coefficient 
and to the uncertainty of the experimental data. Then, we develop
two modification schemes to tune the scattering length
to small positive value using continuous wave laser 
field of experimentally feasible characteristics. In addition, we found that the  scattering length of $^{135}$Cs is positive and thus this isotope  is a suitable candidate for Bose-Einstein condensation of cesium.

The best known potential for the ground triplet state of Cs$_2$
$^3\Sigma_u^+ (6S + 6S)$, is a combination of quantum chemistry calculations by Foucrault {\it et al} \cite{foucrault92} 
fitted at the distance $\approx 20$ Bohr to 
the asymptotic behavior $C_6/R^6$. In the present calculation we used two different values of the coefficient $C_6$ calculated by Marinescu and Dalgarno \cite{marinescu95} and by Derevianko {\it et al} \cite{derevianko99}.
This potential has $55\pm 2$ bound vibrational levels.
The uncertainty of the number is due to the uncertainty of the
potential curves at small distances.
Other potentials used for triplet  Cs$_2$ \cite{kokkelmans98}
lead to a different number of bound vibrational levels, $58\pm 2$.
We checked the influence of the higher order dispersion
terms of the long range potential on the results of the calculations.
Since no significant differences have been found, we kept only the term
$C_6/R^6$ as a representant of the asymptotic behavior.
Using this potential all the bound state wave functions and the threshold
scattering wave functions were calculated using the Mapped Fourier 
Grid Method  \cite{kosloff88,KDKM99,KDKM00}. This approach enable us to
obtain an exact and fully-quantum time-independent solution of 
the Schr\"odinger equation without applying any approximation.
This represent an original and direct approach to the determination
of the scattering length.
A grid of 557 points covering 20000 Bohr was used. Due to the exponential
uniform convergence of the method we were able to converge
the error in phase of the wave functions to an accuracy of $\sim 10^{-6}$.
At the far end of the grid we used either fixed or absorbing boundary conditions.
We found that as expected the threshold continuum wave functions are not 
affected by the choice of boundary conditions.
The accuracy of the calculation is determined by 
the short-range part of the potential
where the inaccuracy of the depth of the well is estimated
to be about 40 $cm^{-1}$ \cite{foucrault92}.
This inaccuracy leads to an asymptotic accumulated phase of more than 
$\pm 2\pi$ thus the scattering length cannot be determined from 
{\it ab initio} calculations.

A binary scattering event is completely determined by the inter-atomic potential
and the boundary conditions. Since the scattering length is determined
by the conditions of vanishing asymptotic kinetic energy it is solea
determined by the potential. For heavy colliders such as Cs$_2$
the scattering length is an extremely sensitive function of the potential.
Its value is related to the accumulated phase $\phi$ of the wave function 
from the inner turning point at zero energy to the infinity. 
The phase $\phi$ for the zero
energy is estimated semiclassically as $\int_{R_i}^{\infty}\sqrt{-2\mu U(R)}dR$ where $R_i$
is the inner turning point of the potential,  $U(R)$ is the potential and
$ \mu$ is the reduced mass of $^{133}$Cs$_2$.
For the potential used for Cs$_2$, $\phi$ is 54.6 $\pi$.
 From the Bohr-Sommerfeld
 quantization condition the number of bound levels should be 54. The number
of levels is obtained by a direct solution of the Schr\"odinger
equation is the same. As 
a consequence, variations of $\pm \pi/2$ will change the scattering length
from negative to positive. These facts pose a 
computation challenge for {\it ab-initio} determination of the scattering length
but they also offer an opportunity for experimental manipulation.

To overcome this problem an independent method
of determining the phase of the wave function has to be found. 
The photoassociation spectrum from the triplet to the
$0_g^- (6S + 6P_{3/2})$ state obtained by Fioretti 
{\it et al} \cite{fioretti98,fioretti99} is used to 
obtain the phase and to adjust the ground state potential.
The spectrum shows oscillating behavior which reflects 
the variation of the Franck-Condon factors of the transition between the 
$^3\Sigma_u^+ (6S + 6S)$ and rovibrational levels of the 
$0_g^-(6S + 6P_{3/2})$ state.
The oscillations  in the spectrum reflect the nodal structure 
of the continuum wave functions in the energy range of  
(200 $\mu$K) \cite{fioretti99}.
Using the Mapped Fourier Grid Method  \cite{kosloff88,KDKM99,KDKM00}, 
the Franck-Condon factors between the 
vibrational wave functions of the $0_g^- (6S + 6P_{3/2})$ state 
and the continuum  wave functions of the $^3\Sigma_u^+ (6S + 6S)$ state
were calculated. 
These factors give relative intensities of the spectral
lines of the $0_g^- (6S + 6P_{3/2})$ spectrum. 
The excited potential curve $0_g^- (6S + 6P_{3/2})$ for short distances
was obtained 
by a diagonalization of the $^3\Sigma_g^+ (6S + 6P)$ 
and  $^3\Pi_g^+ (6S + 6P)$ 
curves of the $0_g^-(6S+6P)$ symmetry while accounting of the 
spin-orbit interaction ($V_{so}(Cs)=$554.1 cm$^{-1}$).  In the asymptotic region
the potential is fitted to the RKR-potential obtained from the experiment
\cite{fioretti99}.

\begin{figure}[h]
\centerline{\psfig{figure=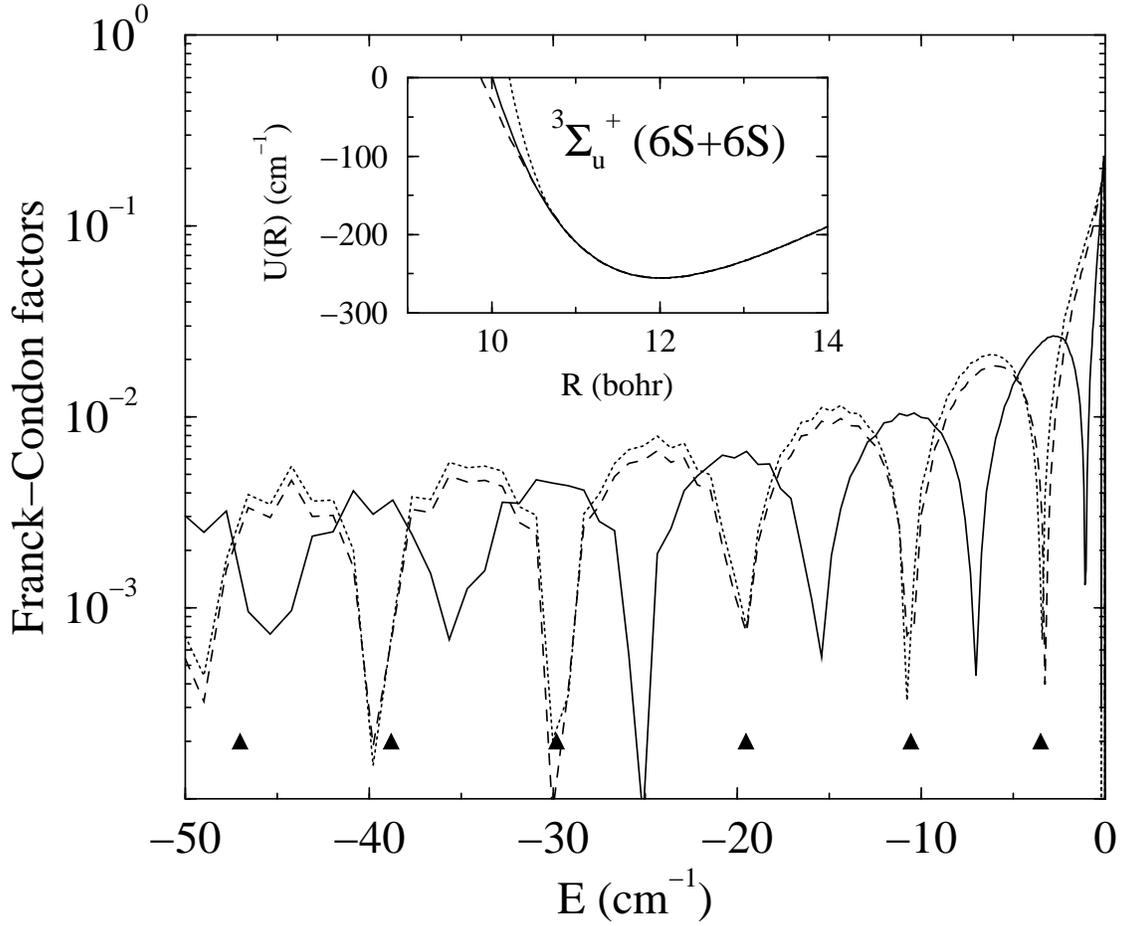,width=0.9\textwidth}}
\caption{The variation of the Franck-Condon factors for 
the transition between the $^3\Sigma_u^+ (6S + 6S)$ 
and the $0_g^- (6S + 6P_{3/2})$  
states for the  original (full line)  and modified potentials  
(dotted and dashed lines) with the detuning $E$.  
Triangles indicate position of the experimental spectral nodes. 
Inset shows also the original (full line) potential and two modified 
potentials (dotted and dashed lines) which both reproduce positions 
of nodes observed experimentally. }
\label{fig:FC}
\end{figure}

The inaccuracy of potential leads to a shift of the position of nodes of the 
calculated spectrum. Fig.  (\ref{fig:FC}) shows 
the calculated intensities (full line) and positions 
of nodes observed in the experiment. The calculated spectrum is almost 
completely out of phase with the experiment.
By slightly modifying the inner part of the potential within
the estimated range of accuracy, the experimental observations can be reconstructed.
The variation in the potential can either add or subtract to the phase 
of the asymptotic part of the wave function.
In both cases, we obtained very good agreement with the experimental nodal 
structure as illustrated in figure \ref{fig:FC}. With this adjustment the number
of bond levels is either 54 or 55. 

In order to see the sensitivity to the inaccuracy of the $C_6$ coefficient we used two different values of $C_6$ calculated by Marinescu and Dalgarno \cite{marinescu95} ($C_6=-6331$ a.u.) and by Derevianko {\it et al} \cite{derevianko99} ($C_6=-6899$ a.u.). As a result the inner part of the fitted potential is slightly different. The figure  \ref{fig:FC} shows the fits for $C_6=-6899$ a.u.

\begin{figure}[h]
\centerline{\psfig{figure=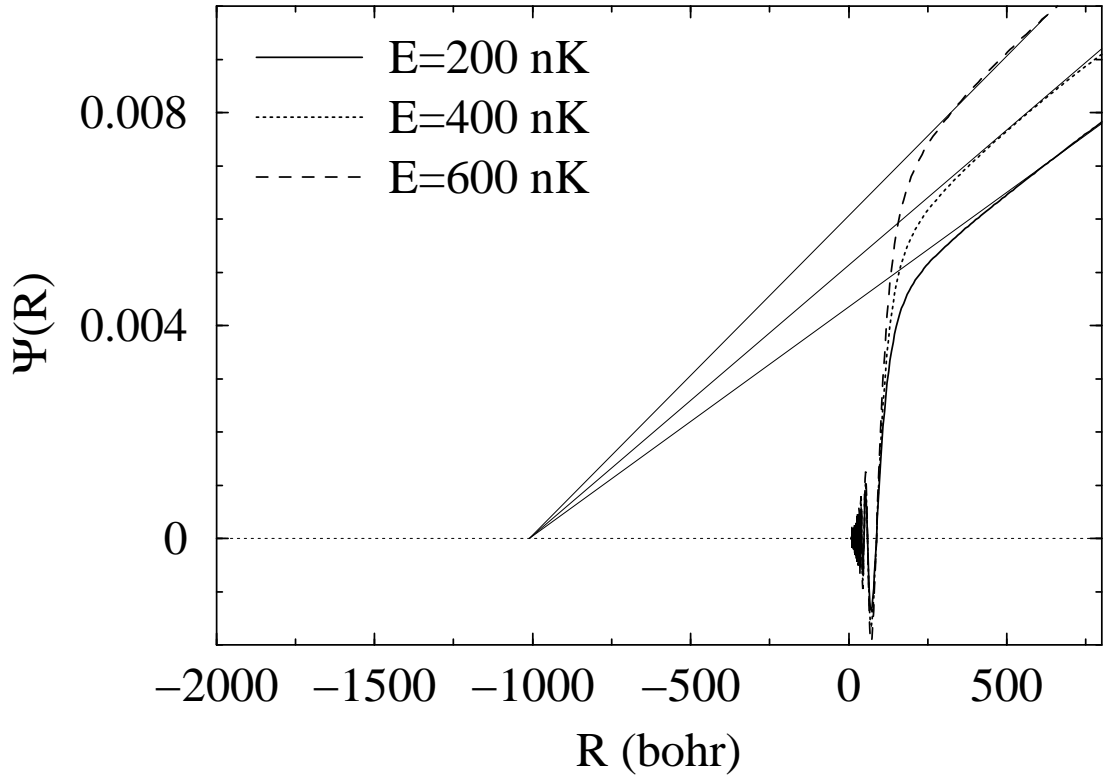,width=0.9\textwidth}}
\caption{The determination of the $^3\Sigma_u^+$ scattering length 
from the continuum wave functions  
slightly above the dissociation limit. The wave functions 
are calculated using the potential adjusted to reproduce 
correctly the oscillations in the photoassociation spectrum.}
\label{fig:wf_fit}
\end{figure}

Using the adjusted  potential which fits  
the experiment the scattering length is recalculated
(Fig. \ref{fig:wf_fit}).
We obtained the value $a_{sc}= -350$ Bohr for $C_6=-6331$ and   $a_{sc}= -950$ Bohr for $C_6=-6899$. For $C_6=-6899$ we made additional calculation in order to see if some inaccuracies could lead to the positive scattering length. First we consider the uncertainty of the position of minima in the experimental spectrum. The fitted potential is changed in such a manner that the calculated minimum prior to the last minimum (the last node is not very pronounced) shifts by 0.2 cm$^{-1}$ in two directions. At this condition the last nodes shifts by 0.1 cm$^{-1}$. The scattering length becomes $a_{sc}= -1040$ Bohr when the nodes shift to the direction of increasing of the binding energy and $a_{sc}= -850$ Bohr for the opposite direction.

We verified also the sensitivity of determination of $a_{sc}$ to the different number of bound levels. With this goal we calculated $a_{sc}$ for the potentials giving 54 and 55 bound levels (dotted and dashed lines on the inset of Fig. \ref{fig:FC} correspondingly). We found that once the two potentials are fitted to the same nodal positions,  the difference in $a_{sc}$ for the two potentials is smaller at least by a factor 10 than the difference due to the uncertainty of the positions of minima. This results is easily explained in a spirit of the quantum defect theory: If the long-range behavior of the potential and positions of last nodes of the last bound wave function are fixed, the short range dynamics (in particular - the total number of nodes of the wave function at small distances) does not influence the scattering properties.

The obtained value of $a_{sc}$ 
is in a good agreement with the previous 
experimental and theoretical results. Arndt {\it et al} \cite{arndt97} 
and Leo {\it et al} \cite{leo98} got the absolute value 
$> 260$ Bohr and $> 600$ Bohr respectively. 
Kokkelmans {\it et al} \cite{kokkelmans98} published the 
value $-315$ Bohr to $-380$ Bohr and Legere {\it et al} \cite{legere98} 
$\approx -400$ Bohr. Recently, Drag {\it et al} \cite{crubellier00} 
analyzed the experimental energy of the last bound  vibrational 
levels and determined the triplet state scattering length of cesium 
from the extrapolated position of the outermost nodes of the bound 
vibrational wave function \cite{crubellier99} obtaining the value ranging 
from $-370$ Bohr down to $-825$ Bohr. The value of $C_6=-6510$ was used in
their calculations. 
In addition, we determined the critical $C_6$ coefficient for which the scattering length becomes infinite. We found that for $C_6=-7000$ it is still negative, but  $C_6=-7050$ is the transition value. For $C_6=-7100$ the scattering length becomes positive. We emphasize the comparison of the presented results 
with those of Drag {\it et al} \cite{crubellier00}, Tab. \ref{tab:comparison}, 
since both approaches analyze the same set of experimental data by 
different methods.

The determined absolute value of the scattering length is so large that small variations of the potential can lead to the change of the sign of $a_{sc}$. For example, a low-intensity laser field coupling the ground molecular state with one of excited states at large distances can modify the scattering length. This possibility was discussed by Kagan {\it et al} \cite{kagan97}.  This sensitivity of the sign of $a_{sc}$ to small perturbations has a crucial importance for the character of behavior at ultra cold temperatures. Below we discuss how $a_{sc}$ can be made small
and positive.

The sensitivity of the asymptotic phase  to the inner part of the potential
is the key to modifying the scattering length. 
The first scheme employs a CW laser field to couple the inner part of the 
$^3\Sigma_u^+ (6S + 6S)$ potential to the $^3\Pi_g (6S + 6P)$ electronic state.
The bound and continuum levels are recalculated on the coupled
two surface potentials. The rotating wave approximation
was used to describe the coupling. The $^3\Pi_g (6S + 6P)$ potential
is taken from \cite{WM}.

\begin{figure}[h]
\centerline{\psfig{figure=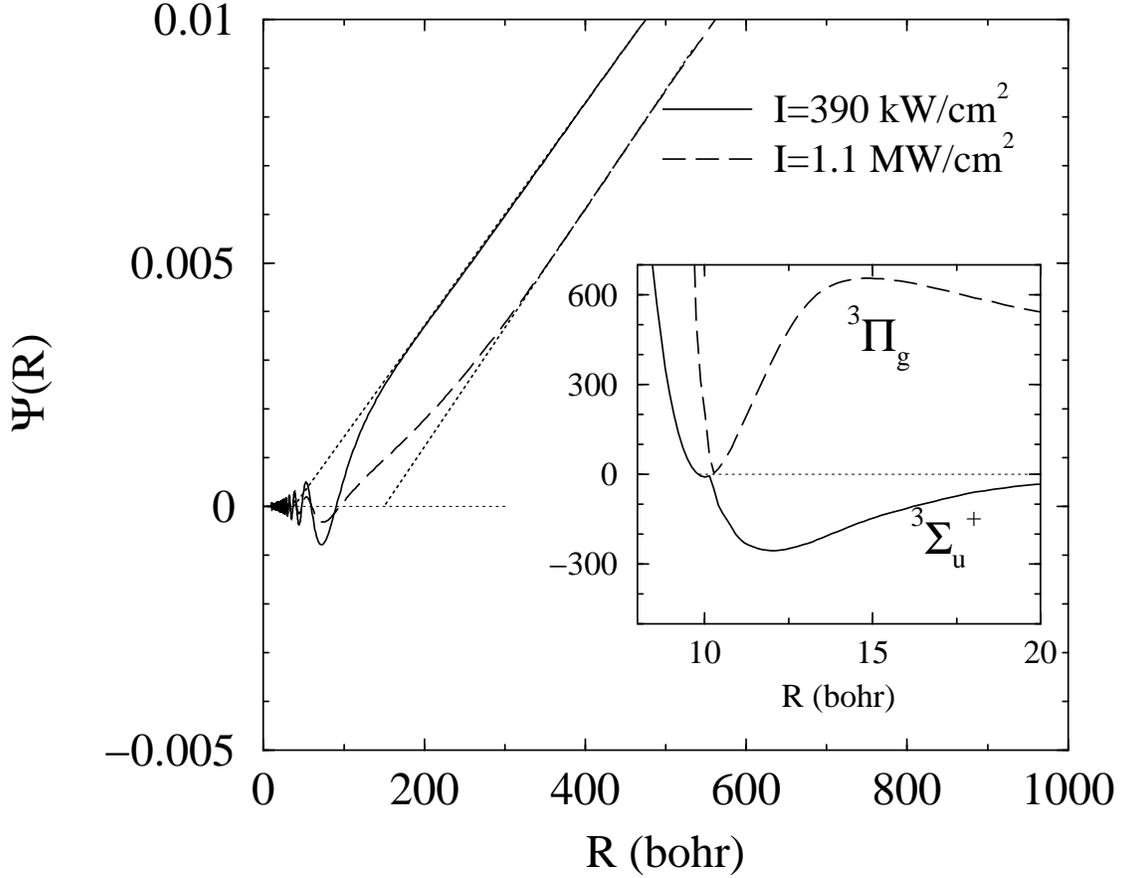,width=0.9\textwidth}}
\caption{The scheme of the modification of the inner part 
of the $^3\Sigma_u^+(6S+6S)$ potential (the inset). 
The two coupled potentials (in cm$^{-1}$) are shown in the 
adiabatic representation using the rotating wave approximation. 
Wave functions with energy $E=0.4\ \mu$K, calculated for two different 
intensities of the laser field. 
Scattering lengths are 30 Bohr for a intensity 
$I=390$ kW and 150 Bohr for $I=1.1$ MW. The energy shift 
$E_f=\hbar \omega$ is the same
for both cases, $E_f=11408$ cm$^{-1}$ - by 324 cm$^{-1}$ red-detuned 
to the $6S\to 6P_{3/2}$ transition.}
\label{fig:strong_field}
\end{figure}

In order to change the scattering length  sufficiently, the adiabatic 
transition probability, $P(^3\Sigma_u^+ \to ^3\Pi_g)$, for 
levels with energy close to the $6S + 6S$ threshold, should be non-negligible. 
For relatively small laser intensity this condition 
is fulfilled only if the potentials  
$^3\Sigma_u^+ (6S + 6S)$ and  $^3\Pi_g (6S + 6P)$ cross at an energy 
close to the $6S + 6S$ dissociation limit. The insert 
in figure \ref{fig:strong_field} shows 
the total two-channel potential in the adiabatic representation. 
The $^3\Pi_g (6S + 6P)$ potential is shifted down by the energy, 
$E_f= \hbar \omega$, determined by the frequency, $\omega$, 
of the field. When the intensity, $I$, of the laser field is low, 
the region of a pseudo-crossing is very narrow. 
Figure \ref{fig:strong_field} shows two 
wave functions obtained for the two different 
intensities for the potentials  shown at the  inset.
We found that with increasing field intensity  
the value of the scattering length gradually 
moves from negative to positive. 
A change of sign  is obtained for the intensity of 300 kW/cm$^2$.

The second scheme,  is based on a resonant coupling of the 
continuum wave functions of the $^3\Sigma_u^+(6S+6S)$ state  
to one of vibrational levels of the $^3\Sigma_g^+ (6S + 6P)$ 
electronic excited state published in \cite{WM}.  
Such a Feshbach resonance results 
in modification of the triplet state scattering length. 
The sign and value of the scattering length depends on the 
position of such a level in respect to the $6S+6S$ threshold.

Employing the Mapped Fourier Grid method, the
low energy continuum wave functions are calculated
for two coupled potentials for different 
intensities of the laser field. The Feshbach resonance is induced by 
the lowest vibrational level of the excited state negatively
detuned from the ground state dissociation threshold
by the energy $\Delta = 90$ MHz.
From the obtained wave functions  the scattering length is determined. 
A gradual increase of the value of 
the scattering length is observed reaching high positive values. 
Since the position (and a width) of the Feshbach resonance depends 
strongly on the coupling, the scattering length can be changed 
employing a much smaller field intensity compared to the off-resonant 
laser manipulation (scheme I).  
The scattering length could be modified from $-350$ Bohr to $300$  Bohr
using a field intensity of only 2.5 kW/cm$^2$. The projected population in 
the excited state is less than $10^{-3}$.

We emphasize that the present method of solution of the
Sch\"odinger equation on the electronic states 
coupled by a field does not rely on any approximative treatment like
perturbation theory and is therefore exact within a given physical model.
That contrasts with  the previous approaches by Fedichev {\it et al} 
\cite{fedichev96} and Bohn \cite{bohn97}.

It is also advantageous to tune the light to one of 
lowest vibrational levels of the whole $6S+6P$ manifold which have
allowed dipole transition.
In this way  one can eliminate the coupling 
with other vibrational levels of other molecular potentials. 
This choice of a vibrational level for the Feshbach 
resonance ensures that a two-photon excitation process 
to the $(6P + 6P)$ manifold is avoided. 
The coupling is also  chosen to occur in the short-range region
where only a very small fraction of the density resides.
This will further reduced loss processes.

The proposed scattering length manipulation techniques employs light 
induced a coupling of the short-range part of molecular electronic potential. 
Therefore, we have to consider the variation of the dipole moment 
due to the molecular orientation with respect to the axis of the field vector. 
We suggest to solve this problem by using three perpendicular 
laser beams with perpendicular polarizations
 for irradiation of the cold cesium sample, which prevents 
any interference effect. In this setup, 
we can assume the maximal variation of the dipole moment projection 
on the field by $\approx 15 \%$. We studied the robustness 
of the positivity of the scattering length by computing 
its value for both proposed schemes using field intensity 
varying by several orders of magnitude. We found that 
both schemes are sufficiently robust to keep the inter-atomic 
interaction repulsive and, hence, to allow the Bose-Einstein condensation. 

Using the adjusted potential the scattering 
properties of the isotopes  $^{135}$Cs and  $^{137}$Cs were calculated
(half-life's are $2.3\cdot 10^6$ and $30.2$ years respectively).   
The scattering length of $a_{sc}=165$ Bohr for $^{135}$Cs was obtained. 
This result is in good agreement with the value $138$ Bohr 
published by \cite{kokkelmans98}. Therefore, the 
$^{135}$Cs isotope is a good candidate for Bose-Einstein condensation. 
For the  $^{137}$Cs, we found  almost zero scattering 
length suggesting that this radioactive isotope 
can behave like a non-interacting Bose gas. In both cases, we tested
the value of the scattering length with respect to the number of bound
levels of the original fitted potential used. 
We found no significant difference; specifically, $\approx 7$ Bohr for
$^{135}$Cs (55 and 56 bound levels)  and $\approx 15$ Bohr for the $^{137}$Cs 
isotope (56 and 57 bound levels).

The spontaneous emission broadening of the Feshbach resonance 
discussed by \cite{fedichev96,bohn97} is eliminated in the present scheme
in two different ways. First, the population transfer at the short range 
part of the potential where the resonance is induced is very low due to 
the small amplitude of the ground state continuum wave function.
That results in a small value of the transition dipole moment
between both electronic states. Since its order of magnitude is $10^{-4}$,
the resulting leading term in the spontaneous emission rate 
$\vert \langle \psi_g \vert \mu \vert \psi_e \rangle \vert ^2$
is approximately $10^{-8}$ extending the spontaneous emission time-scale
to values beyond the microsecond range. That reduces 
the spontaneous emission broadening of the resonance width as well as
it eliminates the related leakage mechanism competing with 
the Bose-Einstein condensation.
Besides, in agreement with Bohn \cite{bohn97},
the coupling is mediated by the field whose intensity,
the Rabi frequency, results in the time-scale much shorter, 
approximately 150 ps, than
the time-scale of the spontaneous emission and, hence, the spontaneous 
emission broadening of the Feshbach resonance is negligible compared
to its field induced width. 

The manipulation of the scattering properties can be 
checked in other  applications. 
For example  photoassociation spectroscopy 
 \cite{stwalley99} can be carried out 
using cold alkali atoms with modified  scattering properties. 
The correspondence between the scattering 
length and the nodal structure of the Franck-Condon factors suggests 
a direct experimental test of the proposed schemes  
employing a  CW field with intensities obtainable by
a diode laser radiation source. 
The technique can be used to  modify existing  experiments done with Rb 
\cite{gabbanini99} or  Cs  \cite{fioretti98,fioretti99}. 

The change of the scattering length 
from  negative to positive results in decreasing the population 
density at short inter-nuclear distance by several orders of magnitude. 
This  will decrease the probability 
for cold molecule formation mediated by three body 
collisions \cite{takekoshi98}. 
As a result the reduction in the
Franck-Condon overlap and transition dipole 
matrix elements will decrease the spontaneous decay 
loss in the cold molecule formation via 
photoassociation \cite{fioretti98,gabbanini99,nikolov99,nikolov00}.

Recently, S\"oding {\it et al} \cite{soding98} measured very high 
inelastic collision rates of the spin flip of ground state 
cesium atoms which results from the negative scattering length. 
The attractive interaction enhances the ground state population 
in the inner part of the potential ruling out the Bose-Einstein 
condensation. Our calculations show that the positive scattering 
length due to the proposed manipulation schemes will reduce the short 
range population at least by a factor of 30. This decreases the 
spin flip scattering rate almost by three orders of magnitude 
removing another obstacle to the the Bose-Einstein 
condensation in cesium.

In conclusion, the present calculation  reproduces the 
oscillations in the $0_g^-(6S + 6P_{3/2})$ photoassociation spectrum of 
$^{133}$Cs by a small adjustment to the $^3\Sigma_u^+(6S+6S)$ potential. 
On this potential we were able to calculate
the scattering length of the $^3\Sigma_u^+(6S+6S)$ state 
finding a  a value $-350$ Bohr. Three different 
schemes to change  the scattering length from negative 
to positive were studied. 
The scattering length can be modified by 
a  laser field either off-resonant with the 
$6S\to 6P$ transition or on resonance with one of lowest vibrational 
levels of $6S+6P$ potentials. 
The third scheme consists in replacing the $^{133}$Cs 
by  the $^{135}$Cs isotope. For this case the scattering length is 165 Bohr. 
We also found that another obstacle to the Bose-Einstein condensation, 
the large rate of inelastic collisions, can be overcome by tuning  
the scattering length.

We are grateful to A. Fioretti for many fruitful discussions.
The work was supported by the Israel Science Foundation (Moked). 
The Fritz Haber Research Center is 
supported by the Minerva Gesellschaft f\"ur die Forschung, GmbH M\"unchen, FRG.

\newpage

\begin{table}[h]
\caption{The scattering length determination.}
\begin{tabular}{|c|c|c|}
\hline
\hline
 & $C_6$ & $a_{sc}$ \\
\hline
\hline
the present work & -6331 & $\approx -350$ \\
\hline
Ref. \cite{crubellier00} & -6510  & -370 to -825 \\
\hline
the present work & -6899 & $\approx -950$ \\
\hline
the present work & -7050 & the transition \\
\hline
\end{tabular}
\label{tab:comparison}
\end{table}

\begin{figure}[h]
\centerline{\psfig{figure=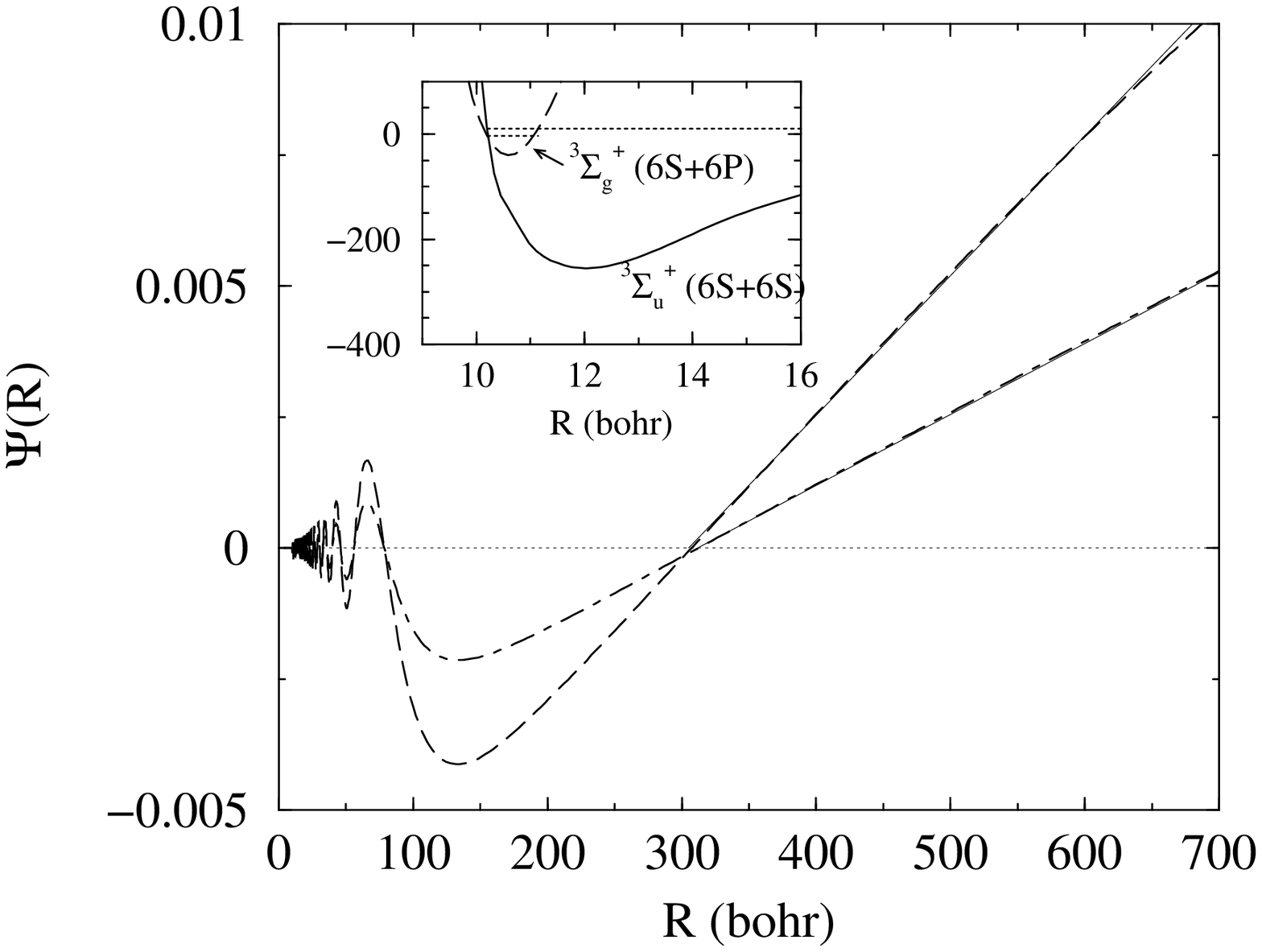,width=0.9\textwidth}}
\caption{The continuum wave functions with the scattering 
length of $300$ Bohr obtained by a resonant coupling of a 
vibrational level in the  $^3\Sigma_g^+ (6S + 6P)$ state to the 
$^3\Sigma_u^+ (6S + 6S)$ potential. The field intensity was  2.5 kW/cm$^2$. 
The inset shows the  position of the  Feshbach resonance 
on the two dressed potentials (energy of the potentials in cm$^{-1}$).
The energy shift $E_f$ is 8545 cm$^{-1}$  - by 3187 cm$^{-1}$ red-detuned 
to the $6S\to 6P_{3/2}$ transition.}
\label{fig:predis}
\end{figure}


\newpage
\begin{thebibliography}{99}
\bibitem{tiesinga92} 
E.~Tiesinga {\it et al.}, Phys. Rev. A {\bf 46}, R1167 (1992).

\bibitem{tiesinga93}
E.~Tiesinga, B.J.~Verhaar, and H.T.C.~Stoof, Phys. Rev. A {\bf 47}, 4114 (1993).

\bibitem{vogels97}
J.M.~Vogels {\it et al.}, Phys. Rev. A {\bf 56}, R1067 (1997).

\bibitem{inouye98}
S.~Inouye  {\it et al.}, Nature (London) {\bf 392},151 (1998).

\bibitem{courteille98}
Ph.~Courteille {\it et al.}, Phys. Rev. Lett. {\bf 81}, 69 (1998).

\bibitem{vuletic99a} 
V.~Vuletic {\it et al.}, Phys. Rev. Lett. {\bf 82}, 1406 (1999).

\bibitem{vuletic99b} 
V.~Vuletic {\it et al.}, Phys. Rev. Lett. {\bf 83}, 943 (1999).

\bibitem{moerdijk96}
A.J.~Moerdijk, B.J.~Verhaar, and T.M.~Nagtegaal, Phys. Rev. A {\bf 53}, 4343 (1996).

\bibitem{marinescu98}
M.~Marinescu and L.~You, Phys. Rev. Lett. {\bf 81}, 4596 (1998).

\bibitem{fedichev96}
P.O.~Fedichev {\it et al.}, Phys. Rev. Lett. {\bf 77}, 2913 (1996).

\bibitem{bohn97}
J.L.~Bohn and P.S.~Julienne, Phys. Rev. A {\bf 56}, 1486 (1997).

\bibitem{kagan97}
Yu.~Kagan, E.L.~Surkov, and G.V.~Schlyapnikov, Phys. Rev. Lett. {\bf 79}, 2604 (1997).

\bibitem{foucrault92} M.~Foucrault, P.~Milli\'e, and J.P.~Daudey, J. Chem. Phys. {\bf 96}, 1257 (1992).

\bibitem{marinescu95} M.~Marinescu and A.~Dalgarno, Phys. Rev. A {\bf 52} 311 (1995).

\bibitem{derevianko99} A.~Derevianko, W.R.~Johnson, M.S.~Safronova, and J.F.~Babb, Phys. Rev. Lett. {\bf 82}, 3589 (1999).

\bibitem{kokkelmans98}  S.J.J.M.F.~Kokkelmans, B.J.~Verhaar, and K.~Gibble Phys. Rev. Lett. {\bf 81}, 951 (1998).

\bibitem{kosloff88}
 R. Kosloff, J. Phys. Chem. {\bf 92}, 2087 (1988).

\bibitem{KDKM99} V.~Kokoouline {\it et al.}, J. Chem. Phys. {\bf 110}, 9865 (1999).

\bibitem{KDKM00} V.~Kokoouline {\it et al.},   Phys. Rev. A (August 2000).

\bibitem{fioretti98}
A.~Fioretti {\it et al.}, Phys. Rev. Lett. {\bf 80}, 4402 (1998).

\bibitem{fioretti99}
A.~Fioretti {\it et al.}, Eur. Phys. J. D {\bf 5}, 389 (1999).


\bibitem{arndt97}  M.~Arndt {\it et al.}, Phys. Rev. Lett. {\bf 79}, 625 (1997).

\bibitem{leo98}  P.J.~Leo {\it et al.}, Phys. Rev. Lett. {\bf 81}, 1389 (1998).

\bibitem{legere98}  R.~Legere and K.~Gibble, Phys. Rev. Lett. {\bf 81}, 5780
(1998).

\bibitem{crubellier00} C.~Drag {\it et al.},  Phys. Rev. Lett. {\bf 85}, 1408 (2000).

\bibitem{crubellier99}
A.~Crubellier {\it et al.},   Eur. Phys. J. D {\bf 6} 211, (1999).

\bibitem{WM} W.~Meyer, (Private communication); N. Spiess, Ph. D thesis, Fachbereich Chemie, Universit\"at Kaiserslautern, (1989).

\bibitem{stwalley99}
W.C.~Stwalley and H.~Wang, J. Mol. Spectr. {\bf 195}, 194 (1999).

\bibitem{gabbanini99}
C.~Gabbanini {\it et al.},  Phys. Rev. Lett. {\bf 84}, 2814 (2000).

\bibitem{takekoshi98}
T.~Takekoshi, B.M.~Patterson, and R.J.~Knize, Phys. Rev. Lett. {\bf 81}, 5105 (1998).

\bibitem{nikolov99}
N.~Nikolov {\it et al.},  Phys. Rev. Lett., {\bf 82}, 703  (1999).

\bibitem{nikolov00}
N.~Nikolov {\it et al.},  Phys. Rev. Lett., {\bf 84}, 246 (2000).


\bibitem{soding98} J.~S\"oding {\it et al.},  Phys. Rev. Lett. {\bf 80}, 1869 (1998).


\end{thebibliography}
\end{document}